\documentclass[aps,pra,twocolumn]{revtex4}
\usepackage{graphicx,color}% Include figure files

\usepackage{bm}% bold math
\usepackage{amssymb}
\usepackage{amsmath}
\usepackage{amstext}
\usepackage{latexsym}

\usepackage[normalem]{ulem}
\begin{document}

\title{Ground state of low dimensional dipolar gases: \\linear and
zigzag chains} \author{G. E. Astrakharchik$^1$, Giovanna
Morigi$^2$, Gabriele De Chiara$^2$, and J. Boronat$^1$}
\affiliation{$^1$ Departament de F\'{\i}sica i Enginyeria Nuclear, Universitat Polit\`ecnica de Catalunya, E-08034 Barcelona, Spain\\
$^2$ Grup d'\`Optica, Departament de F\'{\i}sica, Universitat
Aut{\`o}noma de Barcelona, E-08193 Bellaterra, Spain}
\date{\today} \begin{abstract} We study the ground state phase
diagram of ultracold dipolar gases in highly anisotropic traps.
Starting from a one-dimensional geometry, by ramping down the
transverse confinement along one direction, the gas reaches
various planar distributions of dipoles. At large linear
densities, when the dipolar gas exhibits a crystal-like phase,
critical values of the transverse frequency exist below which the
configuration exhibits novel transverse patterns. These critical
values are found by means of a classical theory, and are in full
agreement with classical Monte Carlo simulations. The study of the
quantum system is performed numerically with Monte Carlo
techniques and shows that the quantum fluctuations smoothen the
transition and make it completely disappear in a gas phase. These
predictions could be experimentally tested and would allow one to
reveal the effect of zero-point motion on self-organized
mesoscopic structures of matter waves, such as the transverse
pattern of the zigzag chain. \end{abstract} \pacs{81.30.Bx,
67.80.-s, 64.70.Tg}
%phase diagrams, 81.30.Bx
%Quantum crystals, 67.80.-s
%Quantum phase transitions, 64.70.Tg
\maketitle

\section{Introduction}

Ultracold atoms and molecules are an attractive playground for
studying fundamental properties of matter. Theoretical and
experimental investigations pursue a fundamental understanding of
the quantum dynamics of matter at ultralow temperatures and
explore applications for quantum technologies, such as quantum
metrology and information
processing~\cite{Maciej-Review,Bloch-Review}. One relevant issue
in this context is the realization and control of
strongly-correlated systems with cold atoms. In this respect,
ultracold dipolar gases play an important role, as the nature of
the dipolar interaction allows one to observe the interplay
between quantum degeneracy and long-range forces. For this reason
they are also interesting candidates for studying statistical
theories of quantum long-range interacting
systems~\cite{Assisi-Book}. Dipolar gases of ultracold atoms have
been experimentally realized with Chromium
atoms~\cite{Pfau07,Pfau08}. In these experiments the strength of
the $s$-wave scattering interaction is conveniently controlled by
tuning the Feshbach resonance and can be made vanishing, thus
leading to the realization of purely dipolar systems. Stability
against collapse is warranted by polarizing the gas in a
two-dimensional geometry, such that the dipolar interactions are
purely repulsive~\cite{Demler07,Grigory07}.

Different phases of dipolar gases of atoms or polar molecules have
been theoretically predicted as a function of density and
dimensionality. In two dimensions, in presence of periodic
potentials the effect of long-range forces gives rise to the
appearance of novel quantum phases~\cite{Menotti07}. In a bulk
system, recent theoretical work predicted the creation of
self-organized structures in two dimensions: the ground state may
exhibit the typical features of a crystal or of a quantum fluid
depending on the density~\cite{Demler07,Grigory07}. In a
one-dimensional geometry, Luttinger liquid models describe the
long-range properties of dipolar gases at ultralow
temperatures~\cite{Citro07}. In this case, by tuning the density,
the phase of the gas undergoes a crossover from a Tonks--Girardeau
gas to a crystal-like phase~\cite{Arkhipov05}. The effect of
transverse quantum correlations in two-dimensional and
one-dimensional systems, arising from the long-range dipolar
force, have been analyzed in stacks of pancake traps~\cite{Wang06}
and in planar arrays of one-dimensional tubes~\cite{Kollath08}.
Quasi-ordered systems of polar molecules in one- and
two-dimensions have been recently considered for quantum
information processing~\cite{DeMille,Rabl07}.

One dimensional systems can be experimentally realized with highly anisotropic traps~\cite{Birkl92,Kinoshita05}. Classically, trapped ultracold dipolar systems may exhibit long-range order in one dimension at vanishing temperatures, such that the equilibrium configuration is solely determined by the interplay between the classical repulsive potential and the external confinement.
In the quantum regime the effect of zero point motion is expected to modify substantially the crystalline properties.This situation must be compared with trapped ultracold ions, interacting via
Coulomb repulsion, where structural phase transitions in
selforganized structures have been extensively
studied~\cite{Birkl92,Fishman08}. Here, the effects of quantum
degeneracy~\cite{Schulz93} are negligible in typical experimental
setups~\cite{Javanainen}, and the structures at experimentally
accessible low temperatures are essentially determined by the
classical potential.

In this article we study theoretically the competition of
long-range interactions and quantum fluctuations in the transverse
stability of a one-dimensional ultracold dipolar gas in a highly
anisotropic trap. We determine the properties of the quantum
ground state using quantum Monte Carlo methods. In particular,
we analyze transverse correlations in the parameter regime in
which the gas is in the quasi-ordered phase, and where the
one-dimensional structure is unstable with respect to increasing
the atomic density and/or to reducing the transverse potential. We
see that transverse patterns are formed, giving rise to mesoscopic quantum structures, where the transverse density distribution exhibits first two peaks, and then, by further
opening the trap, multiple peaks. Differing from ionic Coulomb
crystals, here quantum fluctuations smoothen the transition from
the single- to the double-peaked distribution. Nevertheless, we
observe a relatively sharp transition to the various transverse
structures for parameter regimes which can be identified by means
of a classical theory.

This work is organized as follows. In Sec.~\ref{Sec:2} we
introduce the theoretical model for a two-dimensional gas of
dipolar bosons in presence of tight transverse confinement and
identify the relevant length scales. In Sec.~\ref{Sec:3} we
discuss the phase diagram of the system, which we obtain using a
full quantum numerical simulations; we identify the
parameter regime where the linear-zigzag chain transition could be
observed. In Sec.~\ref{Sec:4} the conclusions are drawn.

\section{Theoretical model} \label{Sec:2}

We consider a system of ultracold bosons (atoms or polar
molecules) of mass $m$ possessing large dipole moments, which are
confined in the  $x-y$ plane, while their dipolar moments are
aligned perpendicularly to the plane by an external field. In this
limit the interaction is the repulsive dipolar potential
$$V(\rho)=\frac{C_{dd}}{4\pi|{\bm \rho}|^3},$$
where $\rho$ is the polar coordinate on the plane and $C_{dd}$ denotes the dipolar
interaction strength. Here, $s$-wave scattering is neglected,
assuming that it vanishes due to a properly tuned Feshbach
resonance, see for instance Ref.~\cite{Pfau08}. The gas is assumed
to be homogeneous along the $x$-direction, where it is
characterized by the linear density $n$, and it is confined along the $y$-direction by a harmonic potential of frequency
$\nu_t$. This external potential sets the characteristic length
$$a_{ho}=\sqrt{\frac{\hbar}{m\nu_t}},$$ determining the transverse size of
the single-particle wavepacket, and which we choose as unit
length. Denoting by ${\bm \varrho}_j={\bm \rho}_j/a_{ho}=(\tilde
x_j,\tilde y_j)$ the rescaled coordinates, the corresponding
dimensionless Hamiltonian $\tilde{H}=H/\hbar\nu_t$ reads
\begin{equation}\label{Ham}
\tilde{H}=\frac{1}{2}\sum_j\left(-\frac{\partial^2}{\partial \tilde  x_j^2}-\frac{\partial^2}{\partial\tilde y_j^2}+\tilde y_j^2+\sum_{i\neq  j}\frac{\tilde{r}_0}{|{\bm \varrho}_i-{\bm \varrho}_j|^3}\right),
\end{equation}
where the parameter $\tilde{r}_0=r_0/a_{ho}$ is the
characteristic length of quantum coherence in dipolar gases,
$$r_0=\frac{mC_{dd}}{4\pi\hbar^2},$$ in units of $a_{ho}$.

The effective one-dimensional Hamiltonian is recovered from
Eq.~(\ref{Ham}) when the chemical potential $\mu$ is much smaller
than the level spacing of the transverse oscillator, $\mu \ll
\hbar\nu_t$. In the quasi-ordered phase, for $nr_0\gg 1$, this
corresponds to the inequality $E_{\rm cr}^{(1D)}/N\ll \hbar\nu_t$,
where $E_{\rm cr}^{(1D)}= N(nr_0)^3\zeta(3)\hbar^2/(mr_0^2)$ is
the potential energy of a classical crystal~\cite{Arkhipov05}, and
which leads to the relation $n\ll r_0^{-1/3}$. In the quantum gas
regime, for $nr_0\ll 1$, the gas is essentially described by a
Tonks--Girardeau gas \cite{Girardeau60} and the condition to be fulfilled is $E_{\rm
TG}^{(1D)}/N=\pi^2\hbar^2n^2/(6m)\ll \hbar\nu_t$, which is
equivalent to the requirement $na_{ho}\ll 1$. The expression  for
$E_{\rm TG}^{(1D)}/N$ does not depend on the strength of the
dipolar interaction, as the interparticle distance is large and
the potential interaction can be neglected. For later convenience,
we introduce the rescaled density $$\tilde{n}=na_{ho},$$ such
that when $\tilde{n}\tilde{r_0}\gg 1$ the system is in the
quasi-ordered phase, while for $\tilde{n}\ll 1 $ it is a
Tonks--Girardeau gas.

\section{Phase diagram} \label{Sec:3}

\begin{figure} \begin{center}
\includegraphics[width=0.7\columnwidth, angle=-90]{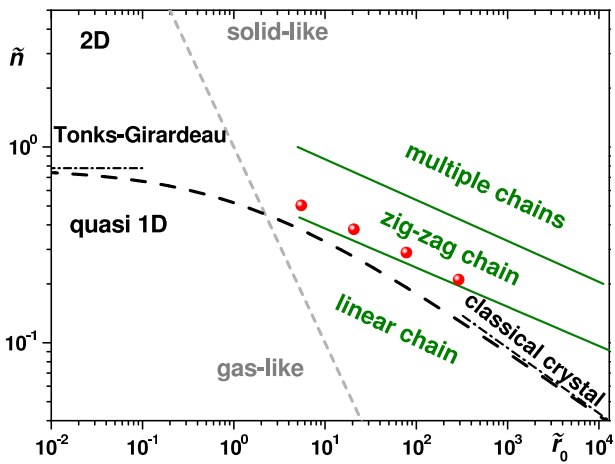}
\caption{(color online) Phase diagram as a function of the
parameters $\tilde{r}_0$ and $\tilde{n}$. The dashed black line
corresponds to the curve $E^{1D}/N=\hbar\nu_t$ and identifies the
two regimes, where the dynamics is essentially one or two
dimensional. The Tonks--Girardeau and the classical crystal limits
are explicitly indicated in the plot (dashed-dotted lines). The
short-dashed grey line, at $\tilde{n}\tilde{r}_0=1$, localizes the
crossover between gas- and solid-like phases. The green solid
lines separate different phases (linear, zigzag, multiple chain)
in the classical system. The red circles are results of quantum
Monte Carlo calculations and correspond to the appearance of a
double-peak structure in the radial density profile (see text and
Fig.~\ref{FigRD}). The size of the symbols denotes the error
bars.} \label{Fig:1} \end{center} \end{figure}

Using the criteria in Sec.~\ref{Sec:2}, we can identify different
phases of the ground state of the dipolar system in a phase
diagram with axes $\tilde r_0$ and $\tilde n$, and which is
reported in Fig.~\ref{Fig:1}. Here, the black dashed line
corresponds to the curve $E^{(1D)}/N=\hbar\nu_t$, where the energy per particle
of the one-dimensional dipolar gas, given in~\cite{Arkhipov05}, is
equal to the level spacing of the transverse confinement
$\hbar\nu_t$. This line separates the two regimes, where the dynamics is
essentially one or two dimensional. The short-dashed line corresponds to
the curve $\tilde n \tilde r_0=1$, which separates the gas from
the quasi-ordered phase. These lines are to be intended as
indicators, the transition of the gas from one phase to another (from one-dimensional to two-dimensional; from gas to quasi-ordered phases) being a crossover. The behavior deep below the black dashed line
of this phase diagram has been studied in Ref.~\cite{Citro07},
where it was shown that the dipolar system is essentially a
Luttinger liquid with Luttinger parameter $K<1$. Although there is
only one phase as the thermodynamic functions are continuous,
nevertheless signatures of quasi order, due to the dipolar
interactions, can be identified in the energy and in the structure
form factor~\cite{Arkhipov05}.

\subsection{Classical theory at $T=0$}

\begin{figure} \begin{center}
\includegraphics[width=0.7\columnwidth, angle=-90]{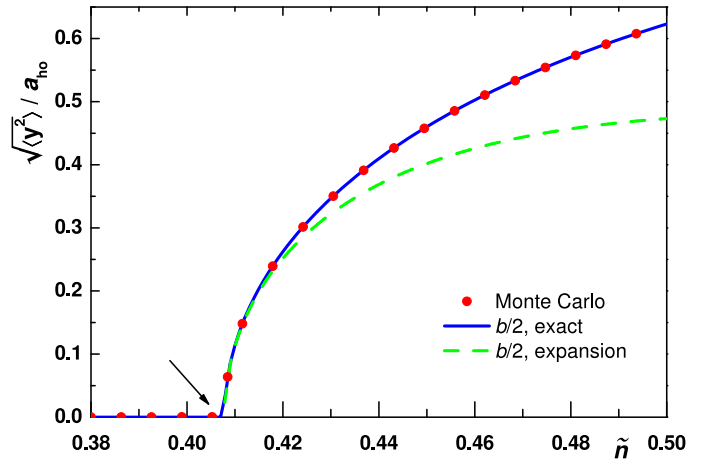}
\caption{(color online) Radial width $\sqrt{\langle y^2\rangle}/a_{ho}$ as a function of $\tilde n$ for the classical dipolar gas at $T=0$ and $\tilde{n}\tilde r_0=3$ close to linear--zigzag transition. The solid line corresponds to the solution of Eq.~(\ref{Eq:b}), which describes the transition from a linear chain to a zigzag configuration, the symbols to the results of a classical Monte Carlo simulation, the dashed line is an expansion close to transition point, Eq.~(\ref{Eq:b1}), the arrow shows the transition point.}
\label{Fig:2a}
\end{center}\end{figure}

We now focus on the quasi-ordered regime, where $\tilde n \tilde
r_0>1$, and first study the dipolar structure by discarding the
kinetic term in the Hamiltonian~(\ref{Ham}). We stress that this approach
gives only a qualitative description, as long-range order cannot
exist in one-dimension at finite temperatures in a dipolar gas,
nor at $T=0$ in a quantum dipolar gas. The predictions of this
classical model are expected to be valid in the asymptotic limit
$\tilde n\tilde r_0\to\infty$ of a quantum system. Keeping this in
mind, we assume that the dipoles are located along the $x$-axis
with interparticle distance $a=1/n$, such that their dimensionless
equilibrium positions are $\tilde x_j=j/\tilde{n}$ and $y_j=0$.
The stability of this configuration requires the linear density
being smaller than a critical value, $\tilde{n}<\tilde{n}_c$,
where $$\tilde{n}_c=\mathcal E \tilde{r}_0^{-1/5}$$ and hence
depends on the trap frequency $\nu_t$, while $\mathcal
E = (8/(93\zeta(5)))^{1/5}=0.6078...$ is a constant. For $\tilde n>\tilde n_c$
the equilibrium positions are distributed in different planar structures,
depending on the value of $\tilde n$. The first structure encountered is a zigzag configuration,
where now $\tilde y_j=(-1)^jb/2$ with $b$ fulfilling the equation
\begin{eqnarray}
&&\sum\limits_{k=0}^\infty \mathcal F_k
b^{2k} = \frac{1}{12(\tilde n\tilde r_0)\tilde{n}^4} \label{Eq:b}\\
&&\mathcal
F_k=\frac{\Gamma(-3/2)\zeta(5+2k)(1-2^{-5-2k})}{\Gamma(1+k)\Gamma(-3/2-k)}\tilde{n}^{2k}
\nonumber \end{eqnarray} These equations are found by evaluating the stable equilibrium points of the potential in Eq.~(\ref{Ham})~\cite{Fishman08}. The expression for the displacement from the axis of the chain close to the critical value $\tilde{n}_c$ reduces to the expression
\begin{eqnarray} b\simeq \frac{\mathcal B}{\tilde{n}_c}
\sqrt{1-\left(\frac{\tilde n_c}{\tilde n}\right)^5},\quad
{\mathcal B}=\sqrt{\frac85\frac{31\zeta(5)}{127\zeta(7)}},
\label{Eq:b1} \end{eqnarray} which provides the critical exponent
of the parameter $b$ for the classical phase transition. At larger
values of $n$ the system exhibits abrupt transitions to more
complex structures. Similar behaviors are also observed in
one-dimensional Wigner crystals~\cite{Birkl92,Piacente04}.
Figure~\ref{Fig:2a} displays the transverse width of the chain
$\sqrt{\langle  y^2\rangle}$ as a function of $\tilde n$ obtained
with a classical Monte Carlo simulation, while the solid line is
the solution of Eq.~(\ref{Eq:b}). The critical value at which the
classical second-order phase transition occurs is indicated by an
arrow. The critical parameters for the linear phase transition in
a classical system are shown in Fig.~\ref{Fig:1} as straight solid
lines.

Thermal and quantum fluctuations modify this behavior, thereby
affecting the form of transverse correlations as a function of the
linear density or of the transverse frequency $\nu_t$. Their
effect is analysed below by means of numerical techniques.

\subsection{Numerical methods}

We resort to Monte Carlo techniques to study numerically the
properties of the dipolar many-body system. The classical system
at finite temperatures is studied by means of a classical Monte
Carlo calculation, where we sample the Boltzmann distribution
$p_{CLS} = \exp\{-E/k_BT\}$ by using the Metropolis algorithm.
The quantum ground-state properties are determined by means of
variational and diffusion Monte Carlo methods~\cite{Boronat94b}.
The trial wave function is chosen in a Bijl-Jastrow form
\begin{equation} \psi_T({\bm \rho}_1, ..., {\bm \rho}_N)=
\prod\limits_{i=1}^N f_1({\bm \rho}_i) \prod\limits_{j<k}^N
f_2(|{\bm \rho}_j-{\bm \rho}_k|), \label{wf} \end{equation}
where the one-body term is $f_1({\bm \rho}) = \exp\{-\alpha
y^2/a_{ho}^2\}$. If the system is in the quasi-one dimensional
regime, where the transverse oscillator is in the ground state,
the variational parameter is the same as for the ground state of
the harmonic oscillator, $\alpha=1/2$. Outside the quasi-one
dimensional regime, the system spreads in the radial direction and
accordingly the value of $\alpha$ is reduced. The two-body term is
chosen as in Ref.~\cite{Grigory07}
\begin{equation} f_2({\bm \rho})\!=\!\left\{ \begin{array}{ll}
C_1 K_0(2\sqrt{r_0/|{\bm \rho}|}), & 0<|{\bm \rho}|<R_{par}\\
C_2\exp\left\{-\frac{C_3}{|{\bm \rho}|}-\frac{C_3}{L_{x}-|{\bm \rho}|}\right\}, & R_{par}\le |{\bm \rho}| < L_x/2\\
1,& L_x/2\le |{\bm \rho}| \end{array} \right., \label{2jastrow}
\end{equation}
where $K_0(|{\bm \rho}|)$ is the modified Bessel function of the second kind, and coefficients $C_1, C_2, C_3$ are fixed by the conditions of the continuity of the wave function and its first derivative. The parameter $R_{par}$ is free, it varies in the interval $0<R_{par}<L_x/2$ and is optimized by a  variational procedure. When the distance between two particles is small, the influence of other particles can be neglected and $f_2(\rho)$ is well approximated by the solution of the two-body scattering problem (see short distance behavior of (\ref{2jastrow})). In this way the divergent behavior of the interaction potential does not cause any numerical instability.
At large distances the two-body term is written in symmetric way.
This ensures that $f'(L_x/2)=0$ and periodic boundary conditions
are properly satisfied.

The results of classical simulations are obtained using an annealing procedure. This means that the classical simulation is started at a large temperature, $k_BT/\hbar\nu_t\approx 1$, and then the temperature is gradually reduced to very small values, where the classical system is frozen. By starting the simulation at a high temperature one is able to explore more effectively the phase space and this helps to avoid the system finding a local minimum instead of a global one. We repeat the annealing procedure several times for each set of parameters $\tilde n$ and $\tilde{r}_0$ thus choosing the lowest energy.

\subsection{Numerical results}

\begin{figure} \begin{center}
\includegraphics[width=0.7\columnwidth, angle=-90]{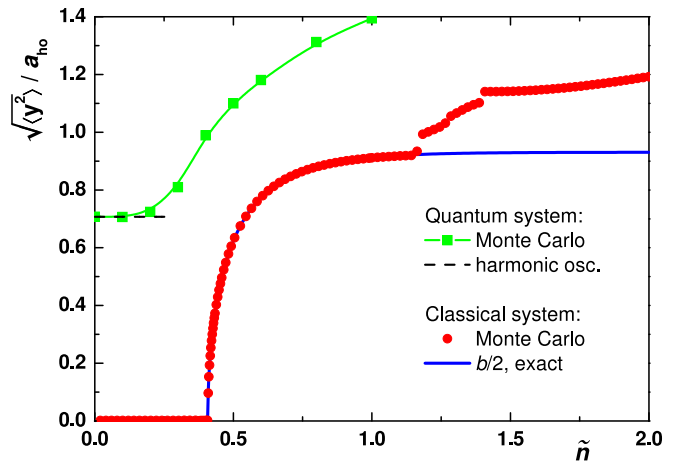}
\caption{(color online) Radial width $\sqrt{\langle y^2\rangle}/a_{ho}$ as a function of $\tilde n$ for $\tilde{n}\tilde r_0=3$ in the classical and quantum dipolar gases at $T=0$.
% Circles correspond to the results of a classical Monte Carlo simulation, for $\tilde n < 1.2$ ({\it i.e.} for linear and zigzag configurations) they coincide with the solution of Eq.~(\ref{Eq:b}), shown as a blue solid line.
Circles correspond to the results of a classical Monte Carlo simulation; blue solid line corresponds to the solution of Eq.~(\ref{Eq:b}). For $\tilde n < 1.2$ ({\it i.e.} for linear and zigzag configurations) both results coincide.
A sudden transition to another, broader, planar structure is observed at larger values of $\tilde n > 1.2$. Squares (connecting line is drawn as a guide to the eye) results of quantum Monte Carlo calculations, dashed line corresponds to the radial width of the transverse harmonic oscillator, $\sqrt{\langle y^2\rangle} = a_{ho}/\sqrt{2}$.}
\label{Fig:2b} \end{center}
\end{figure}

In the quantum system we study the transition from a linear to a
planar configuration by considering the radial spreading $\langle
y^2\rangle$ as a function of density and trap frequency. In
particular, when the chain is in the ground state of the
transverse harmonic oscillator, then $\langle y^2\rangle =
a_{ho}^2/2$, while it becomes larger as the transverse potential
is excited. In Fig.~\ref{Fig:2b} the radial spreading for the
quantum mechanical case is shown as a function of $\tilde{n}$ and
compared with the classical result at $T=0$. Quantum fluctuations
clearly smoothen the transition, and in particular introduce a basic uncertainty related to the oscillator length $a_{ho}$.
Similarly, the transition is smoothened in the classical system at finite temperatures.
It is interesting to notice that the second value,
at which one observes another discontinuity of the classical
radial spreading, is a transition to multiple chains and it is
expected to be of first order~\cite{Piacente04}. The quantum
mechanical behavior at this point will be object of future
studies. The results of the quantum MC simulations corresponds to
the circles in Fig.~\ref{Fig:1} and have been determined by the critical value of the parameters for which two maxima are formed in the radial density profile. We note that the position of the transition
in a quantum system is close to the predictions of a classical
theory with the agreement being even better as $\tilde{r}_0$ is
increased. On the opposite, for small values of $\tilde{r}_0$ the
two-peak structure, characteristic for a zigzag chain, disappears
due to an increased role of quantum fluctuations and the crystal
gets completely melted.

\begin{figure}
\begin{center}
\includegraphics[width=0.7\columnwidth, angle=-90]{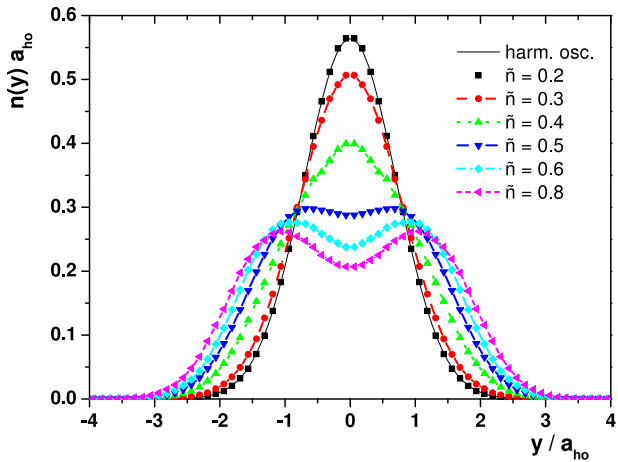}
\caption{(color online) Radial density profile $n(y)$ for fixed $\tilde n  \tilde r_0 = 3$ and at $\tilde{n}=0.2,0.4,0.5,0.6,0.8$ (decreasing the height at $y=0$). This choice of the parameters corresponds to effectively changing the frequency of the harmonic confinement. The radial confinement is normalized to unity, $\int_{-\infty}^\infty n(y)\;dy = 1$. The solid line corresponds to the single-particle wavepacket of the transverse harmonic oscillator and is plotted for comparison. For these parameters, the value at which the classical phase transition occurs is $\tilde n_c\simeq 0.408$.}
\label{FigRD}
\end{center}
\end{figure}

\begin{figure} \begin{center}
\includegraphics[width=0.8\columnwidth,angle=0]{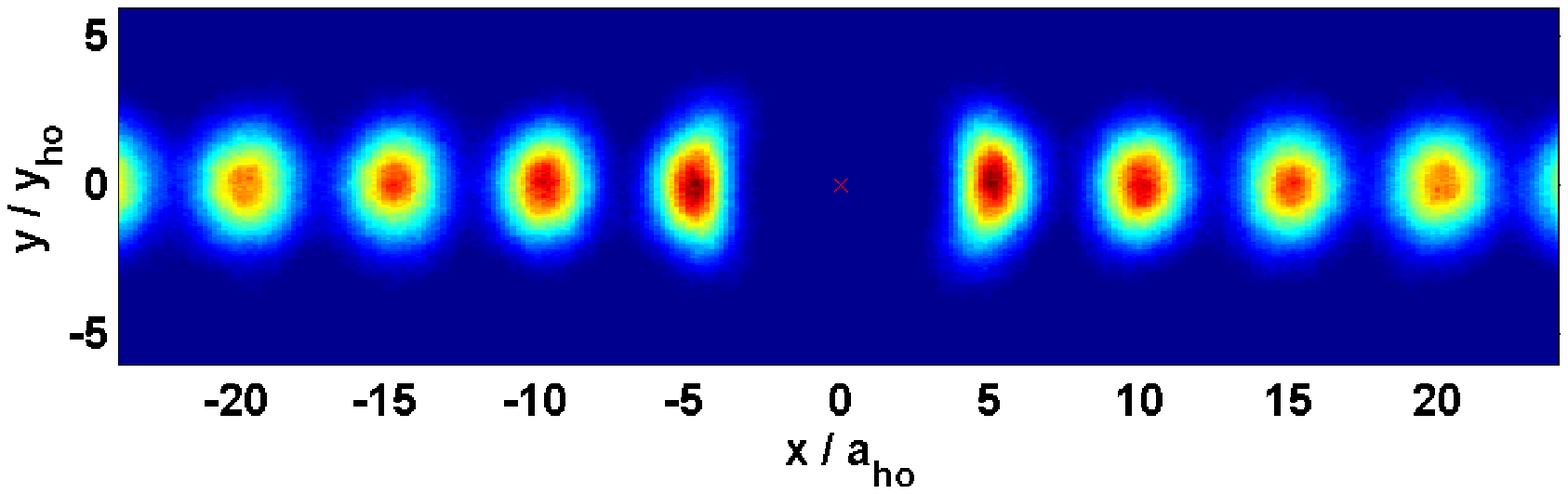}
\includegraphics[width=0.8\columnwidth,angle=0]{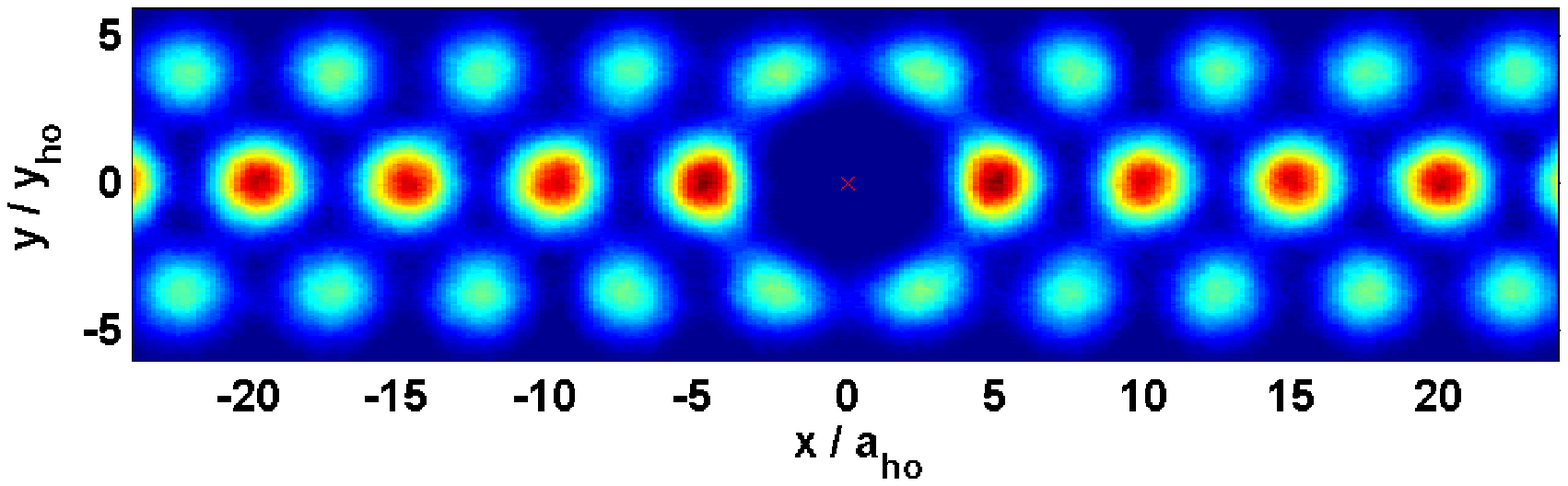}
\caption{(color online) Two dimensional contour plot of pair distribution
function $\langle n({\bf 0})n(x,y)\rangle$ in a quantum system for
fixed value of one-dimensional density $\tilde{n}\tilde r_0 = 40$.
Bright colors correspond to high values of the pair distribution function.
Upper plot, $\tilde r_0 = 100$ (linear chain); lower plot, $\tilde
r_0 = 200$ (zigzag chain).} \label{FigPD} \end{center}
\end{figure}

Figure~\ref{FigRD} displays the radial density profile at fixed
$nr_0$ for various values of $\tilde{n}$, which are distributed
across the value $\tilde n_c=0.408\ldots$ at which the classical
phase transition linear--zigzag occurs for the chosen value of
$nr_0$. For $\tilde{n}<\tilde{n}_c$ the radial density profile is
a single-peaked curve, a Gaussian of width $a_{ho}$, centered at
$y=0$. This is the situation we expect for the quantum linear chain. For $\tilde{n}>\tilde{n}_c$, the density distribution becomes double
peaked, the peaks being symmetrical about $y=0$. The density in
the center is however significantly different from zero: this
effect originates from zero-point quantum fluctuations, which
prevent the particles from being localized at the minima of the
potential energy, which for $\tilde{n}>\tilde{n}_c$ is a double
well potential. At larger values of $nr_0$, nevertheless, one
recovers the expected zigzag configuration for values of
$\tilde{n}$ closer to at $\tilde{n}_c$: in this regime the size of
zigzag spreading is comparable to the oscillator length. The size
of the region, where quantum fluctuations are relevant, shrinks as
the interaction strength, here represented by the parameter
$\tilde{r}_0$, is increased. These planar structures are of
mesoscopic nature as the number of chains is small. Here, the
radial density profile cannot be described by the local density
approximation, which, instead, is generally applicable in a
macroscopic system. Figure~\ref{FigPD} displays characteristic
examples of the pair-correlation function, showing that for
sufficiently large values of the interaction a zigzag distribution
of dipoles is observed.

Thermal fluctuations in the classical systems give rise to a
qualitatively similar behavior. The curves here obtained for the
quantum system are qualitatively reproduced by using a classical
system at the effective temperature $T_{\rm
eff}=\hbar\nu_t/2\kappa_B$.

\subsection{Discussion}

The phase diagram in Fig.~\ref{Fig:1} can be experimentally
explored by changing the density $\tilde n$, thereby moving
vertically, and by varying the transverse frequency $\nu_t$,
thereby moving parallel to the straight short-dashed line. Typical
parameters of 1D experiments~\cite{Kinoshita05}, with $a_{ho}
\approx 35$ nm and $60$ --- $400$ atoms per tube of length $15$
--- $50\mu$m, are $\tilde r_0 \approx 0.07$, $\tilde n \approx
0.04$ --- $0.3$. While $^{52}$Cr has the largest value of $r_0 =
2.4$ nm among all atom species with which condensation has been
reached~\cite{Pfau08}, the use of polar molecules such as CO,
ND$_3$, HCN, CsCl with $r_0 = 5$~nm --- 340~$\mu$m would permit
one to cover regions of the phase diagram up to the classical
region ($\tilde n\tilde r_0\gg 1$).

The formation of the zigzag structure can be experimentally revealed in the structure form factor. The latter, in fact, exhibits an additional peak at the $y$-component of the wave vector $k_y=2\pi/b$, where $b$ is the distance between the two chains forming the zigzag structure.

An important question is whether the system under consideration is superfluid or not. The problem of superfluidity is very delicate when treating (quasi) one-dimensional systems. While different definitions of the superfluidity coincide in homogeneous two- and three- dimensional systems, they lead to contradictive conclusions when applied to a one dimensional system. Indeed, the excitation spectrum of one-dimensional Luttinger liquids touches zero for a non-zero value of momentum, $k = 2\pi n$. Thus, such systems are always normal from the point of view of Landau criterion. Alternatively, one can use the winding number technique\cite{Ceperley} to calculate the superfluid fraction. Applied to the exactly known ground state wave function (for example Calogero--Sutherland system) it predicts a completely superfluid system. In the mean-field limit a system with short range interactions is well described by the Gross--Pitaevskii equation. It is natural to think that in this regime, where the description of a pure condensate applies, the system is superfluid.

A possible way out, proposed in Ref.~\cite{LevPetrovich}, is based on the calculation of the energy dissipation caused by dragging a small probe through the system. The conclusion is that there is always some dissipation, which ranges from negligibly small (in Gross--Pitaevskii limit) to the same as in a normal system (in Tonks--Girardeau limit). There is a continuous crossover from (quasi)-superfluid to normal system. In our case, the limit of small density $\tilde n\to 0$ corresponds to Tonks--Girardeau regime, where the system is normal. Indeed, in this regime the wave function of the system can be mapped onto a wave function of an ideal normal Fermi gas.
%As the density $\tilde n$ is increased, the potential energy of the dipolar interaction becomes important
%. As the interaction is repulsive, its effect would be to increase further energy losses.
%is increased , the potential energy of the dipolar interaction becomes important
%increasing further energy losses.
At larger values of $\tilde n$ the dipolar system corresponds to a Luttinger liquid with even stronger interactions than in the Tonks--Girardeau regime.
As a result our system remains normal for all densities.

\section{Conclusions} \label{Sec:4}

To conclude, we have studied the ground state of a dipolar gas of bosons at $T=0$ in low dimensions starting from quasi one-dimensional  geometry and opening the radial trap, so that a two-dimensional structure develops. 
We do an analytical study of the system in the classical limit close to the transition to the zigzag configuration by applying Landau theory and compare these predictions with numerical simulations using classical Monte Carlo methods. 
We determine the phase diagram numerically using quantum Monte Carlo methods, and study in detail the effect of quantum fluctuations on the linear- zigzag transition. The transition from a one-dimensional to a planar configuration occurs with the creation of a mesoscopic structure in the transverse direction,
which exhibits the main features of the transition from a single to a double and then a multiple chain distribution of dipoles, while quasi-order is observed also in the transverse pair correlation function at large densities. Such patterns are characterized by non-local correlations, which arise from zero-point fluctuations, and which may be important resources for the realization of quantum simulators~\cite{Maciej-Review,Zoller08}. Moreover, the control of atomic patterns has potential applications for nanostructuring processes~\cite{RefLithographyChromium}.

\begin{acknowledgements}

We acknowledge discussions with E.~Demler and Sh.~Fishman. Support
by the ESF (EUROQUAM "CMMC"), the European Commission (EMALI,
MRTN-CT-2006-035369; SCALA, Contract No.\ 015714) and the Spanish
Ministerio de Educaci\'on y Ciencia (Consolider Ingenio 2010
``QOIT''; FIS2005-04181; FIS2007-66944; Ramon-y-Cajal; Juan de la Cierva) are
acknowledged.

\end{acknowledgements}

%\bibliography{dipolechain}

\begin{thebibliography}{99}

\bibitem{Maciej-Review}
%Ultracold atomic gases in optical lattices: mimicking condensed matter physics and beyond
%Authors: Maciej Lewenstein, Anna Sanpera, Veronica Ahufinger, Bogdan Damski, Aditi Sen De, Ujjwal Sen
M. Lewenstein, {\it et al.}, Adv. Phys. {\bf 56}, 243 (2007).

\bibitem{Bloch-Review}
%Many-Body Physics with Ultracold Gases
I. Bloch, J. Dalibard, and W. Zwerger, Rev. Mod. Phys.
\textbf{80}, 885 (2008).

\bibitem{Assisi-Book} A. Campa, A. Giansanti, G. Morigi, and F.
Sylos-Labini, editors, {\it Dynamics and thermodynamics of systems
with long-range interactions: Theory and Experiments}, AIP
Conference Proceedings, vol. 970 (Melville, 2008).

\bibitem{Pfau07}
%title = {Strong dipolar effects in a quantum ferrofluid},
T. Lahaye, {\it et al.}, Nature {\bf 448}, 672 (2007).

\bibitem{Pfau08}
%"Stabilizing a purely dipolar quantum gas against collapse"
T. Koch, {\it et al.}, Nature Physics {\bf 4}, 218 (2008).

\bibitem{Grigory07}
%Quantum Phase Transition in a Two-Dimensional System of Dipoles
G. E. Astrakharchik, J. Boronat, I.L. Kurbakov, and Yu.E. Lozovik,
Phys. Rev. Lett. {\bf 98}, 060405 (2007).

\bibitem{Demler07}
%Strongly Correlated 2D Quantum Phases with Cold Polar Molecules: Controlling the Shape of the Interaction Potential
H.P. B{\"uchler}, {\it et al.}, Phys. Rev. Lett. {\bf 98}, 060404
(2007).

\bibitem{Menotti07}
%Quantum Phases of Dipolar Bosons in Optical Lattices
K. Goral, L. Santos, and M. Lewenstein, Phys. Rev. Lett. {\bf 88}, 170406 (2002);
%Metastable States of a Gas of Dipolar Bosons in a 2D Optical Lattice
C. Menotti, C. Trefzger, and M. Lewenstein,
Phys. Rev. Lett. {\bf 98}, 235301 (2007).

\bibitem{Citro07}
%Evidence of Luttinger-liquid behavior in one-dimensional dipolar quantum gases},
R. Citro, E. Orignac, S. De Palo, and M. L. Chiofalo, Phys. Rev. A
{\bf 75}, 051602(R) (2007).

\bibitem{Arkhipov05}
%Ground-state properties of a one-dimensional system of dipoles
A.S. Arkhipov, G.E. Astrakharchik, A.V. Belikov, and Yu.E. Lozovik, Pis'ma Zh. Eksp. Teor. Fiz. {\bf 82}, 41 (2005).

\bibitem{Wang06}
%Quantum Fluids of Self-Assembled Chains of Polar Molecules
D.-W. Wang, M. D. Lukin, and E. Demler, Phys. Rev. Lett.
\textbf{97}, 180413 (2006)

\bibitem{Kollath08}
%Dipolar Bosons in a Planar Array of One-Dimensional Tubes
C. Kollath, J. S. Meyer, and T. Giamarchi, Phys. Rev. Lett.
\textbf{100}, 130403 (2008).

\bibitem{DeMille} A. Andr\'e, {\it et al.}, Nature physics {\bf
2}, 636 (2006).

\bibitem{Rabl07}
%Molecular dipolar crystals as high-fidelity quantum memory for hybrid quantum computing
P. Rabl and P. Zoller, Phys. Rev. A {\bf 76}, 042308 (2007).

\bibitem{Kinoshita05} T. Kinoshita , T. Wenger, and D. S. Weiss,
%Local Pair Correlations in One-Dimensional Bose Gases
Phys. Rev. Lett. {\bf 95}, 190406 (2005).

\bibitem{Birkl92}
%Multiple-shell structures of laser-cooled 24Mg+ ions in a quadrupole storage ring
G. Birkl, S. Kassner, and H. Walther, Nature {\bf 357}, 310
(1992).

\bibitem{Fishman08}
%Structural phase transitions in low-dimensional ion crystals
S. Fishman, G. De Chiara, T. Calarco, and G. Morigi, Phys. Rev. B {\bf 77}, 064111 (2008).

\bibitem{Schulz93}
%Wigner crystal in one dimension
H. J. Schulz, Phys. Rev. Lett. {\bf 71}, 1864 (1993).

\bibitem{Javanainen}
J.~Yin and J.~Javanainen, Phys.\ Rev.\ A {\bf
51}, 3959 (1995).

\bibitem{Girardeau60}
M. Girardeau, J. Math. Phys. {\bf 1}, 516 (1960).

\bibitem{Piacente04}
%Generic properties of a quasi-one-dimensional classical Wigner crystal
G. Piacente, I. V. Schweigert, J. J. Betouras, and F. M. Peeters, Phys. Rev. B {\bf 69}, 045324 (2004).

\bibitem{Boronat94b}
See, for instance, J. Boronat and J. Casulleras, Phys. Rev. B {\bf 49}, 8920 (1994).

\bibitem{Zoller08} G. Pupillo, {\it et al.}, Phys. Rev. Lett. {\bf
100}, 050402 (2008).

\bibitem{Ceperley} E.L. Pollock and D.M. Ceperley, Phys. Rev. B {\bf 36}, 8343 (1987).

\bibitem{LevPetrovich} G. E. Astrakharchik and L. P. Pitaevskii, Phys. Rev. A {\bf 70}, 013608 (2004).

\bibitem{RefLithographyChromium}
M. Oberthaler and T. Pfau, J. Phys.: Condens. Matter {\bf 15}, R233 (2003).

%\bibitem{Werner05}
%J. Werner {\it et al.} Phys. Rev. Lett. {\bf 94}, 183201 (2005)


\end{thebibliography}
\end{document}